\documentclass[a4paper,11pt]{article}
\usepackage{pos}

\usepackage{comment}
\usepackage{dsfont}
\usepackage{wrapfig}
\usepackage{multicol}

\definecolor{lcolor}{rgb}{0.5,0,0}
\definecolor{citcolor}{rgb}{0,0.3,0.0}
\definecolor{ao(english)}{rgb}{0.0, 0.5, 0.0}

\definecolor{applegreen}{rgb}{0.55, 0.71, 0.0}
\definecolor{cadetblue}{rgb}{0.37, 0.62, 0.63}
\definecolor{cadet}{rgb}{0.33, 0.41, 0.47}
\definecolor{byzantine}{rgb}{0.74, 0.2, 0.64}
\definecolor{orange}{rgb}{1.0, 0.5, 0.0}

\def\bea{\begin{eqnarray}}
\def\eea{\end{eqnarray}}

\def\be{\begin{equation}}
\def\ee{\end{equation}}

\def\vspSmall{\vspace*{0.2\baselineskip}}

\newcommand{\mbf}{\mathbf}

\newcommand{\tpert}{t}
\newcommand{\tcent}{\bar{t}}			
\newcommand{\mrm}{\mathrm}

\newcommand{\Q}{Q}
\newcommand{\wplas}{\omega_{\mrm{pl}}}

\newcommand{\SU}{\mrm{SU}}

\newcommand{\fig}{Fig.~}

\newcommand{\eq}{Eq.~}

\newcommand{\re}{Ref.~}
\newcommand{\res}{Refs.~}

\newcommand{\pToFigs}{.}

\newcommand{\alert}[1]{{\color{red}{#1}}}

\newcommand{\citSmall}{\scriptsize \color{gray}}

\renewcommand{\mathbf}{\vec}

\title{Nonperturbative excitations in overoccupied gluon plasmas}

\author*[a]{Kirill Boguslavski} 
\affiliation[a]{Institute for Theoretical Physics, Technische Universit\"{a}t Wien,\\
1040 Vienna, Austria}

\emailAdd{kirill.boguslavski@tuwien.ac.at}

\abstract{
Motivated by the early-time dynamics of the quark-gluon plasma in high-energy heavy-ion collisions, we extract gluonic spectral functions of overoccupied gauge theories far from equilibrium using classical-statistical lattice simulations and linear response theory. In 3+1 dimensions we find that the spectral function exhibits quasiparticle excitations at all momenta that are mostly consistent with perturbative hard-thermal loop predictions, while partially showing nonperturbative deviations \cite{Boguslavski:2018beu}. In contrast, the structure of excitations in 2+1 dimensions is nontrivial and nonperturbative \cite{Boguslavski:2021buh,Boguslavski:2019fsb}. These nonperturbative interactions lead to broad excitation peaks in the spectral function, demonstrating the absence of soft quasiparticles in these theories. This also suggests that there may be significant nonperturbative corrections present in systems with large momentum anisotropy, which are relevant to phenomenological applications in heavy-ion collisions. 
}

\FullConference{%
 The 38th International Symposium on Lattice Field Theory, LATTICE2021
  26th-30th July, 2021
  Zoom/Gather@Massachusetts Institute of Technology
}

\begin{document}
\maketitle

\section{Introduction}

Motivated by heavy-ion collision experiments, our goal is to study microscopic properties of QCD out of equilibrium from first principles, and to assess their impact on the description of the quark-gluon plasma. This can be done efficiently by studying the spectral functions $\rho(\omega,p)$ of gluons or quarks since they encode the full single-particle excitation spectrum.

Having the initial stage of heavy-ion collisions in mind \cite{Gelis:2010nm,Lappi:2006fp}, we use a weak-coupling ($g^2 \ll 1$) far from equilibrium approach where we study a highly occupied gluon plasma 
with a large gluon distribution function $f(p) \sim 1/g^2$ for soft momenta below a characteristic scale $p \lesssim Q$. In this case both nonperturbative lattice and perturbative diagrammatic methods are available. Focusing on the former, we run classical-statistical lattice simulations in real time and extract the spectral functions. Our results are then compared to perturbative hard-thermal loop (HTL) calculations extended to out-of-equilibrium situations to assess which properties of the spectral functions are beyond HTL results. 
Our main questions involve: 
\begin{itemize}
 \item[$\star$] What are the spectral functions in heavy-ion collisions at early times?
 \item[$\star$] Do they have general features that are common in other non-equilibrium states or in thermal equilibrium at high temperatures?
 \item[$\star$] Are nonperturbative extensions of kinetic theory or anisotropic HTL necessary for a faithful description of the dynamics?
\end{itemize}
To address these questions, we extract spectral functions in a linear-response framework in classical-statistical simulations. Here we apply our framework in Minkowski space-time to isotropic 3+1 dimensional plasmas and (effectively) 2+1 dimensional plasmas, corresponding to the limit of extreme momentum anisotropy. Details of our approach and results can be found in \res\cite{Boguslavski:2021buh,Boguslavski:2019fsb,Boguslavski:2018beu}. Planned extensions of these studies are outlined in the Conclusion.

    
\section{Method \& setup}

We consider $\SU(N_c)$ Yang-Mills theory with classical action 
\be
\label{eq:YM_S}
    S_{\mrm YM}[A] = -\frac{1}{4}\,\int d^{d+1}x\;F_a^{\mu\nu}[A] F^a_{\mu\nu}[A]
\ee
in $d$ spatial dimensions. Simulations are performed for $N_c = 2$ in a gauge-covariant formulation with link fields $U_j(x) \approx \exp\left( ig\, a_s A_j(x) \right)$ discretized on a cubic lattice of size $N_s^3$ with lattice spacing $a_s$. 

Motivated by different initial stages in heavy-ion collisions, we study three models for highly occupied gluonic systems (here in a Minkowski space-time for simplicity): {\em isotropic 3+1 dimensional} system (i.e., $d=3$ in \eqref{eq:YM_S}), a purely {\em 2+1 dimensional} plasma ($d=2$), and a {\em Glasma-like 2+1 dimensional} system. The latter extends the {\em 2+1D} model by including an adjoint scalar field $\phi_a$, and corresponds to a $3+1D$ system with independence of the $x^3$ coordinate. 

In each case we initialize the system with large occupation numbers (with $E = \partial_t A$)
    \begin{align}
f(t{=}0, p \lesssim \Q) \sim \frac{1}{g^2} \gg 1\;, \qquad f(t,p) = \frac{\langle |E_T(t,p)|^2\rangle}{p}\;.
    \end{align}



\subsection{Classical-statistical real-time lattice simulations}

For classical-statistical lattice simulations as employed in our studies, we first set the initial conditions by choosing fields randomly that satisfy $\langle E_T^*(0,\mbf p) E_T(0,\mbf q) \rangle \propto p\,f(0,p)\, (2\pi)^3\delta(\mbf p - \mbf q)$ with transverse fields $E_T^j p_j = 0$ and vanishing $\langle E_L^*(0,\mbf p) E_L(0,\mbf q) \rangle = 0$. We set the gauge fields $A$ similarly, and restore the Gauss law initially if needed. 
Each of the initial configurations $\{ U(0,\mbf x), E(0,\mbf x) \}$ is evolved by solving the classical field equations in a leapfrog scheme in temporal gauge $A_0 = 0$, i.e., $U_0 = \mathds{1}$. 
An observable $O[U,E]$ that depends on the fields is computed at a time $t>0$ by averaging over the initial configurations
\be
O(t) = \frac{1}{\# k} \sum_k O[U(t),E(t)]\,.
\ee



\subsection{Self-similar turbulent attractors}

\begin{figure}[t]
 \centering
 \includegraphics[width=0.48\textwidth]{\pToFigs/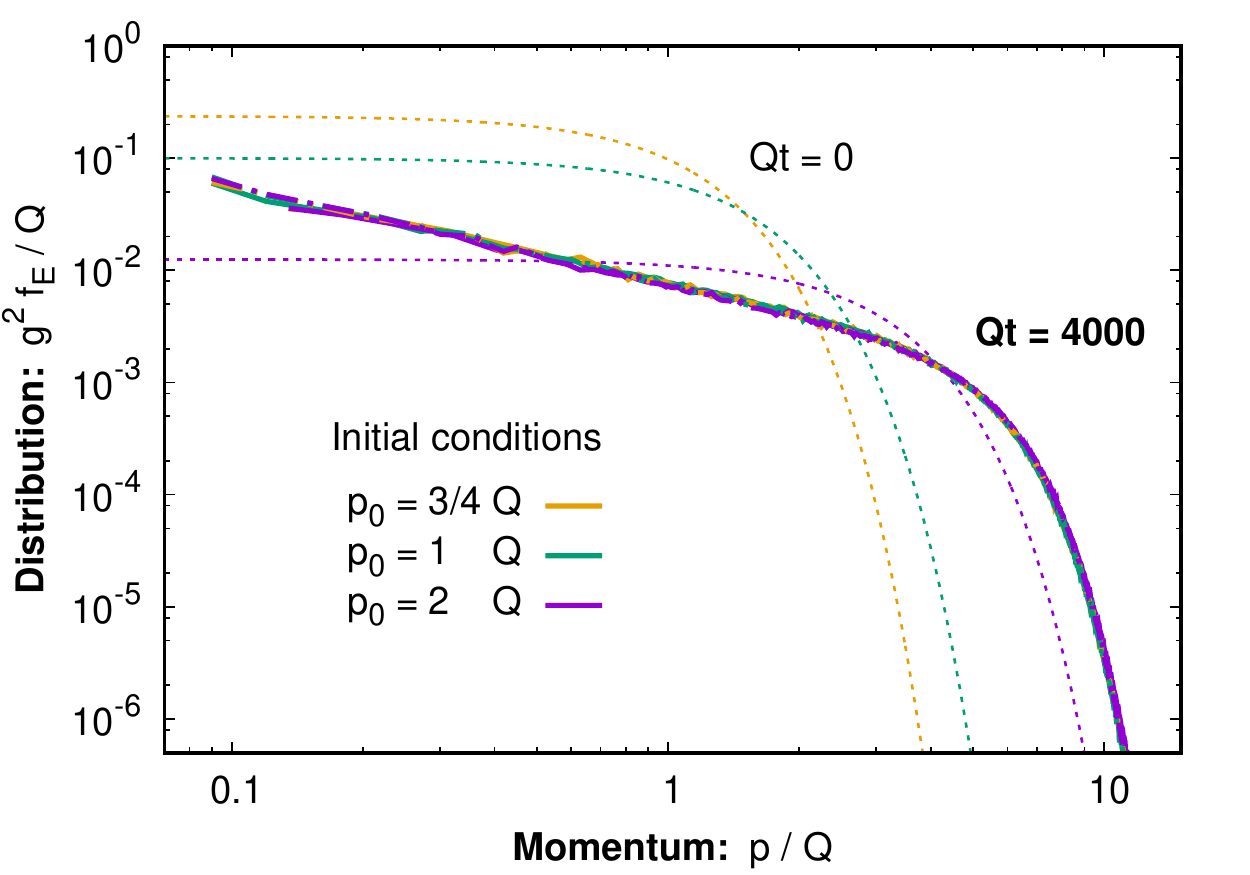}
 \includegraphics[width=0.48\textwidth]{\pToFigs/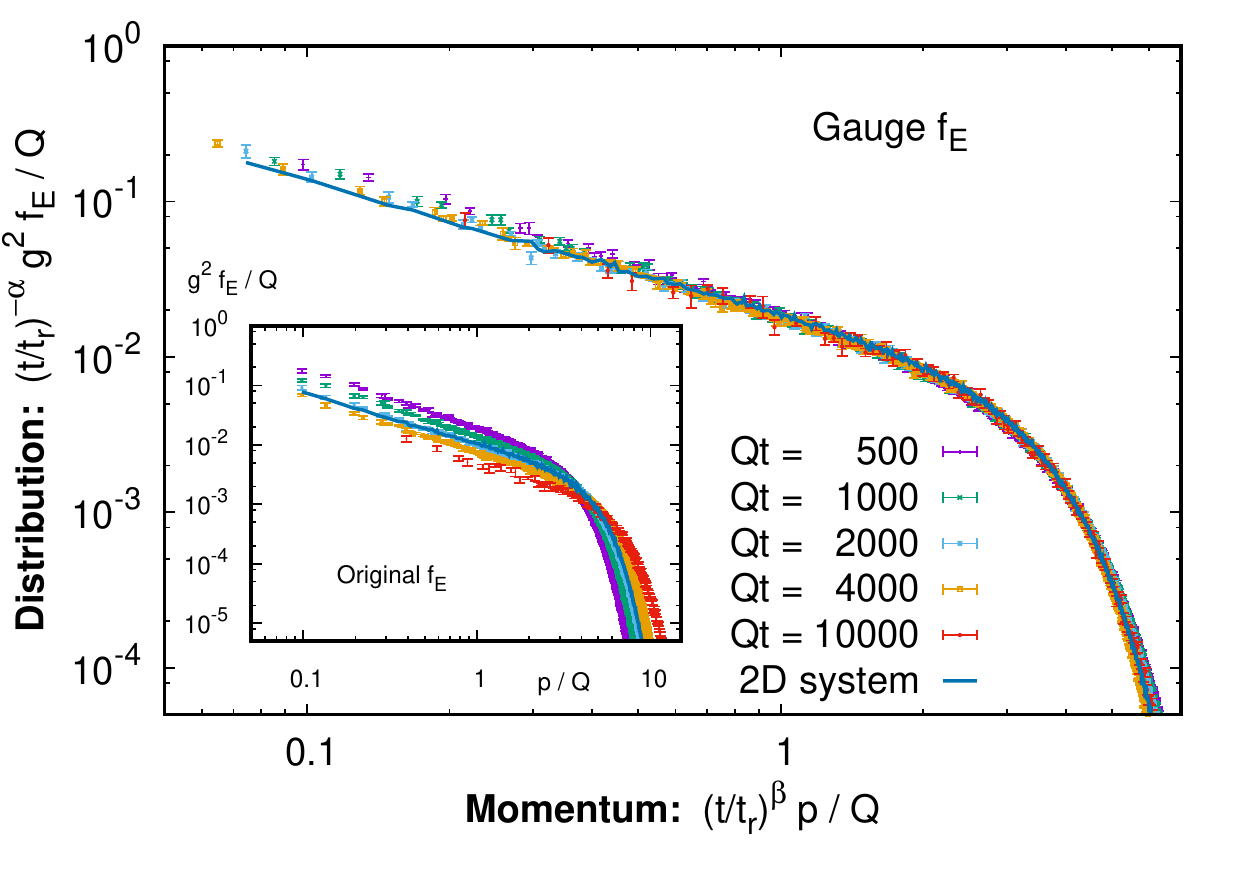}
 \caption{
 Gluon distribution function $f_E \equiv f$ as a function of momentum of a Glasma-like plasma. 
 {\em Left:} Different initial conditions lead to the same attractor (shown at $Q t = 4000$). 
 {\em Right:} Rescaling $f$ and $p$ with universal power-laws of time 
 signals self-similarity. Figures taken from \re\cite{Boguslavski:2019fsb}.
 }
 \label{fig:2D_selfsim}
\end{figure}

Each of the highly occupied systems approaches a self-similar attractor with defining property
\begin{align}
 \label{eq:selfsim}
 f(t,p) = (\Q t)^{\alpha} f_s\left((\Q t)^{\beta} p \right).
\end{align}
The scaling exponents $\alpha$, $\beta$ and scaling function $f_s(p)$ are universal, insensitive to details of initial conditions, and given by
\bea
 \label{eq:selfsim_expo_2D}
 \beta = -1/5\;, &\quad& \alpha = 3\beta \qquad \text{for 2+1D and Glasma-like \cite{Boguslavski:2019fsb}} \\
 \beta = -1/7\;, &\quad& \alpha = 4\beta \qquad \text{for isotropic 3+1D \cite{Berges:2008mr,Kurkela:2012hp}.}
\eea
This is shown in \fig\ref{fig:2D_selfsim} for the example of 2+1D Glasma-like simulations. In the left panel, different initial conditions ($Qt = 0$) lead to the same universal distribution at a later time ($Qt = 4000$). In the right panel, distributions at different times fall on top of each other after rescaling with power-laws of time with the universal exponents in \eqref{eq:selfsim_expo_2D}.


\subsection{Spectral and statistical correlation functions}

Due to \eq\eqref{eq:selfsim}, the evolution of self-similar attractors is known. We extract the spectral and statistical correlation functions at these attractor states because their scaling properties allow us to study the dependence of the correlators on time and momentum scales. For instance, if a quantity in the spectral function scales like a mass $\sim Q(Qt)^{\beta}$, it is likely that the soft scale plays an important role. We will return to this below. 


Hiding indices, the {\em spectral} function is defined as the commutator of fields
\be
 \rho (x', x) = \frac{i}{N_c^2-1}\left\langle \left[ \hat{A}(x'), \hat{A}(x) \right] \right\rangle
\ee
and we also denote $\dot{\rho} = \partial_t \rho$ with $t \equiv x^0$. The {\em statistical} correlation function $\langle EE \rangle$ ($\equiv \ddot{F}$) is in general, out of equilibrium, independent of $\dot{\rho}$ and defined as the anti-commutator of fields
\be
 \langle EE \rangle (x', x) = \frac{1}{2(N_c^2-1)}\left\langle \left\{ \hat{E}(x'), \hat{E}(x) \right\} \right\rangle \,.
\ee
To extract the correlation functions from our simulations, we first fix the residual gauge freedom to Coulomb-type gauge $\left. \partial^j A_j\right|_{t} = 0$ at time $t$ and Fourier transform the correlations in $t'-t$ and $\mbf x' - \mbf x$ to frequency $\omega$ and momentum $\mbf p$. Instead of doing this at fixed $\tcent = \frac{1}{2}(t+t')$, we keep $t$ fixed, which requires the approximation $t \approx \tcent \gg \Delta t \equiv t'-t$ that is satisfied in our simulations. 
The statistical correlator can be measured as
\be
 \langle EE \rangle (t',t,p) = \frac{1}{N_c^2-1}\left\langle E(t',\mbf p)E^*(t,\mbf p) \right\rangle\,.
\ee


\subsection{Nonperturbative computation of spectral function $\rho$}

\begin{wrapfigure}{l}{0.37\linewidth}
\includegraphics[width=0.35\textwidth]{\pToFigs/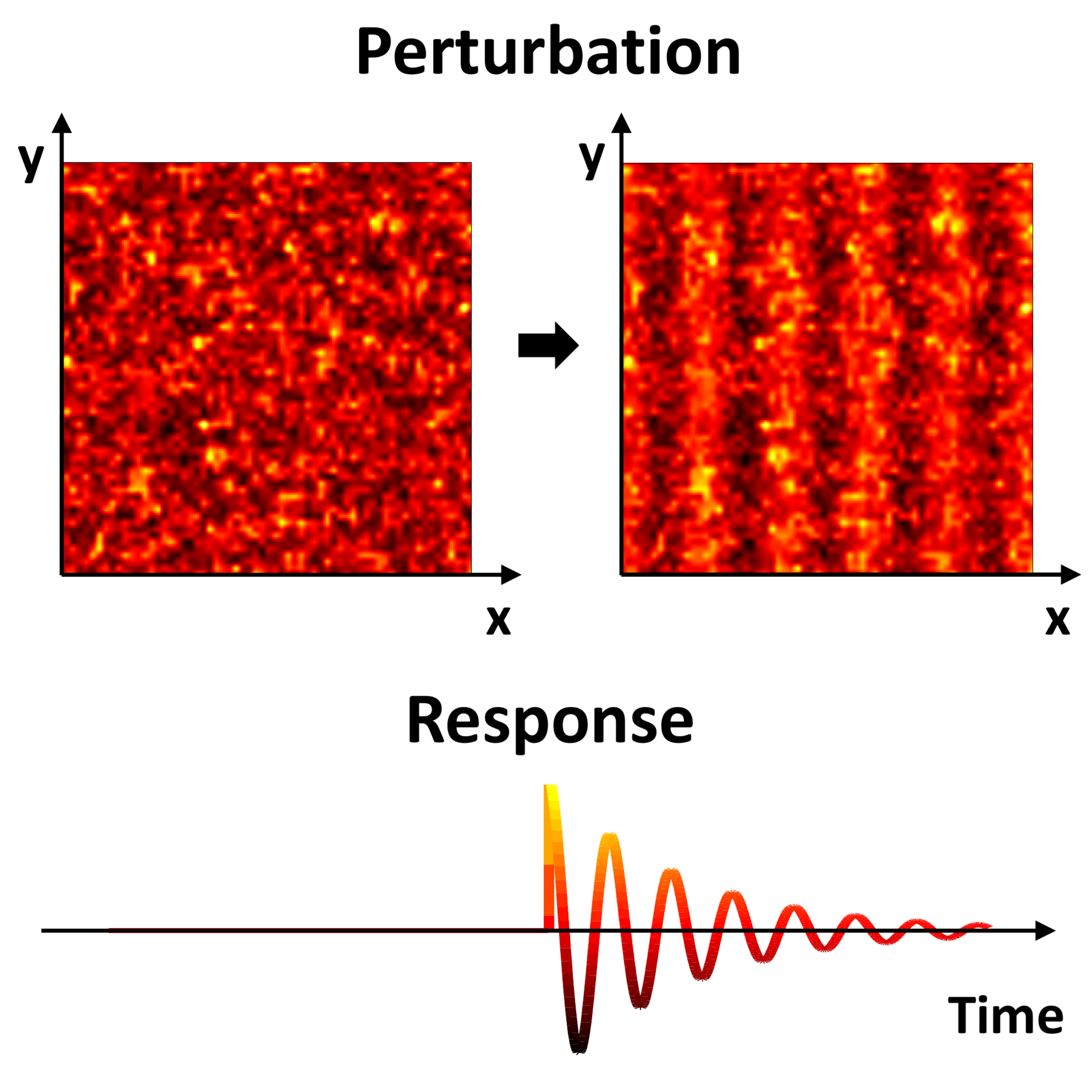}
\caption{Visualization of the method to compute $\rho$ (see \cite{Boguslavski:2018beu}).}
\label{fig:visual_perturb}
\end{wrapfigure}

In order to measure the spectral function, we combine classical-statistical simulations with linear response theory, as introduced in \re\cite{Boguslavski:2018beu} and visualized in \fig\ref{fig:visual_perturb} on the left. 

At time $t$ we perturb the fields with an instantaneous source $j(t',\mbf p) = j_{0}(\mbf p)\, \delta\left( t' - \tpert \right)$. Splitting $A(t,\mbf x) \mapsto A(t,\mbf x) + \delta A(t,\mbf x)$ due to the perturbation, the background field $A$ keeps following the classical Yang-Mills equations while for the linear response field $\delta A$ we solve linearized equations of motion formulated in a gauge-covariant form in \re\cite{Kurkela:2016mhu} that are consistent with Gauss law conservation by construction. 
The solution is linked to the retarded propagator $\langle \delta A(t', \mbf p)\rangle = G_{R}(t', t, \mbf p)\, j_{0}(\mbf p)$ and to the spectral function via $\theta(t' - t) \rho(t',t,p) = G_{R}(t',t,p)$. 

While the spectral function in thermal equilibrium is usually computed by using the statistical correlation function and relating it to the spectral function via the fluctuation dissipation relation (FDR) \cite{Aarts:2001yx,Schlichting:2019tbr}, a separate and independent computation of the spectral function like ours is crucial out of equilibrium since the existence of a FDR is not guaranteed. 
Similar methods to the one for gauge theories described here have also been developed for scalar systems far from equilibrium \cite{PineiroOrioli:2018hst,Boguslavski:2019ecc} and recently introduced for fermionic spectral functions in \re\cite{Boguslavski:2021kdd}. There the Dirac field and the Dirac equation play the role of the linear response field and linearized equation of motion, respectively. 




\section{Numerically extracted spectral functions}

\subsection{Gluon spectral function in isotropic 3+1D plasmas}

\begin{figure}[t]
\centering
\includegraphics[width=0.45\textwidth]{\pToFigs/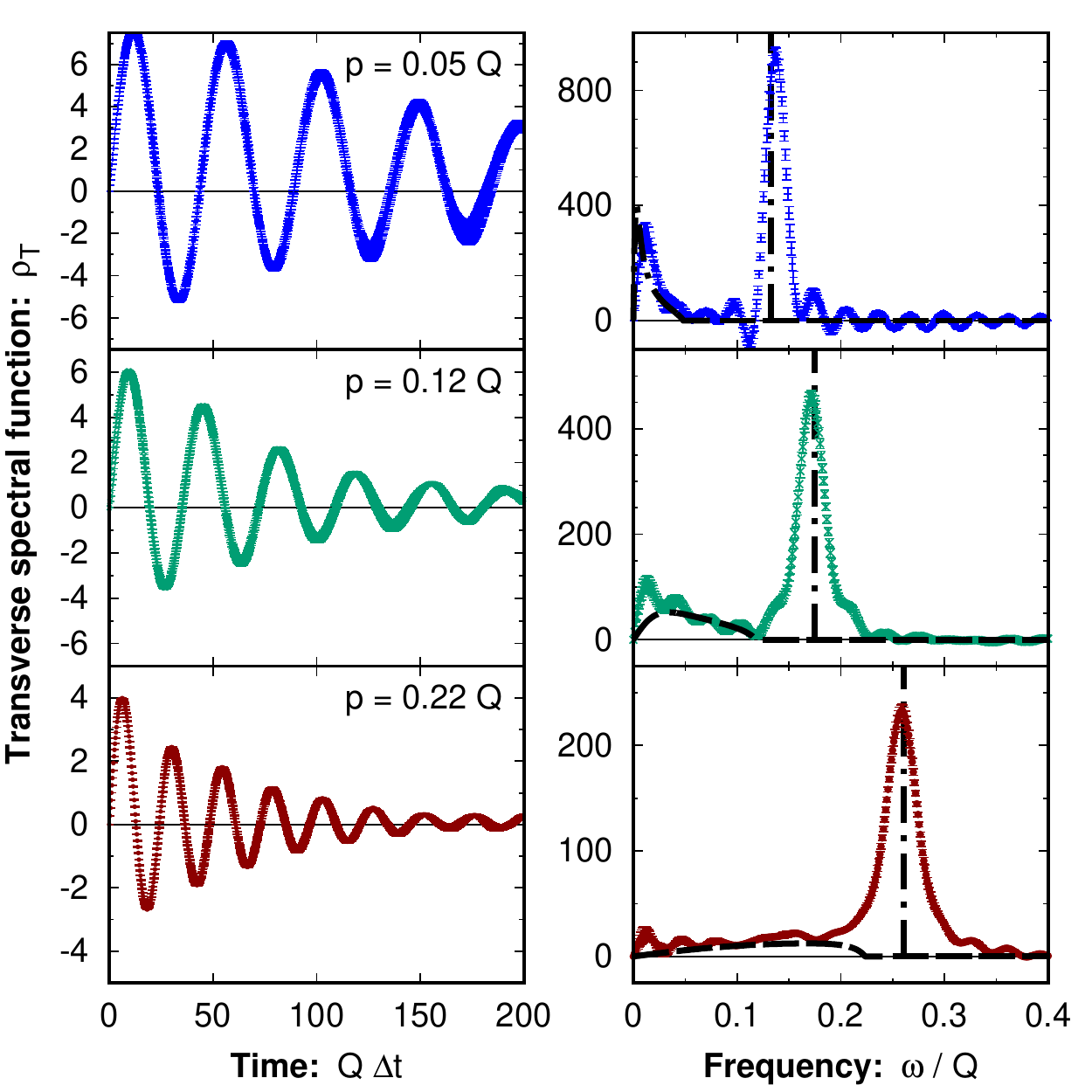}
\includegraphics[width=0.45\textwidth]{\pToFigs/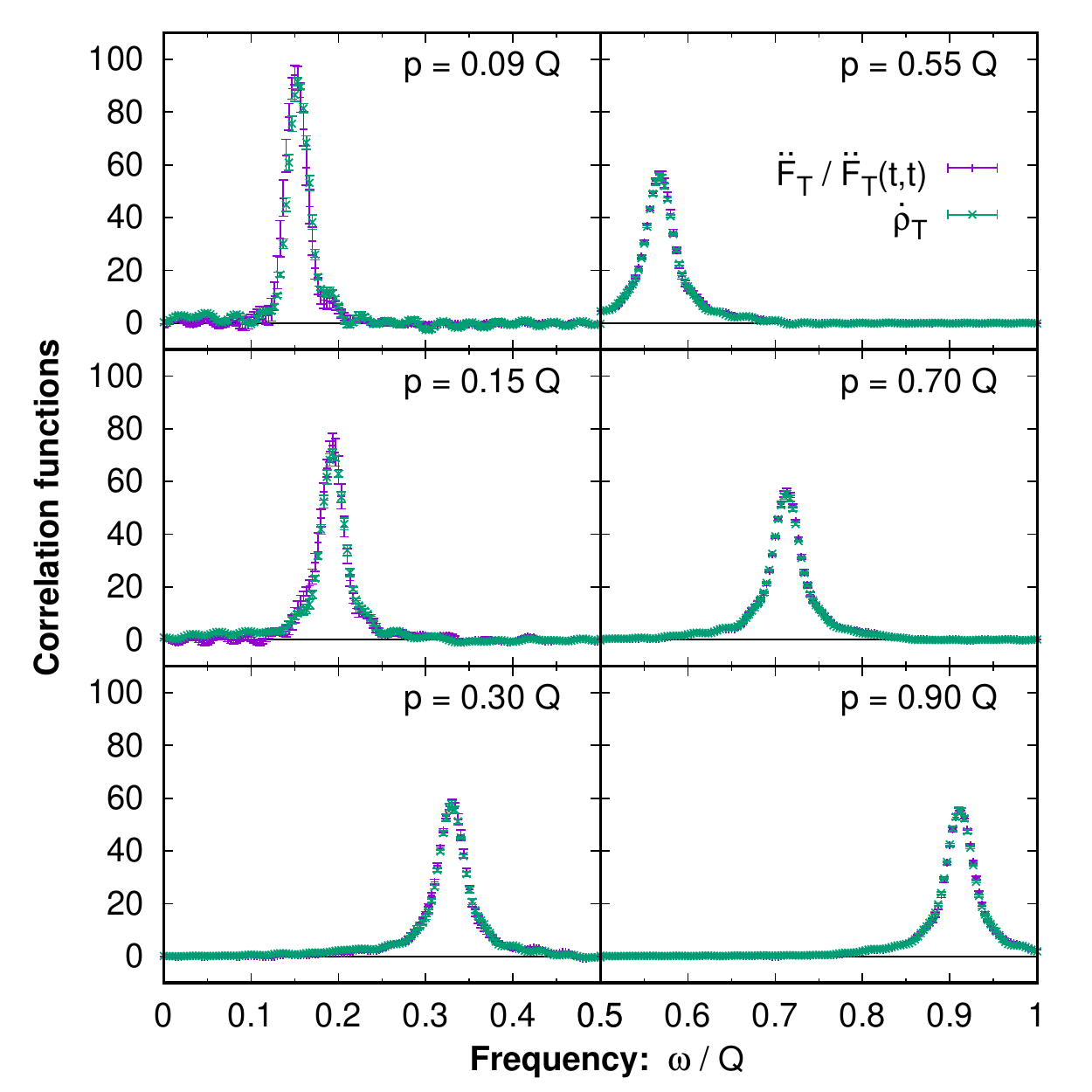}
\caption{Correlations in isotropic 3+1D systems. 
{\em Left:} Spectral function of transversely polarized gluons $\rho_T(\omega,p)$ for different momenta in relative time $\Delta t$ and frequency domains. 
{\em Right:} Normalized correlations $\ddot{F}_T/\ddot{F}_T(t,t) \equiv \langle E_T E_T\rangle/\langle E_T(t) E_T(t)\rangle$ and $\dot{\rho}_T$ for different momenta. Figures taken from \re\cite{Boguslavski:2018beu}.
}
\label{fig:rho_3D}
\end{figure}

\begin{figure}
 \centering
 \includegraphics[width=0.45\textwidth]{\pToFigs/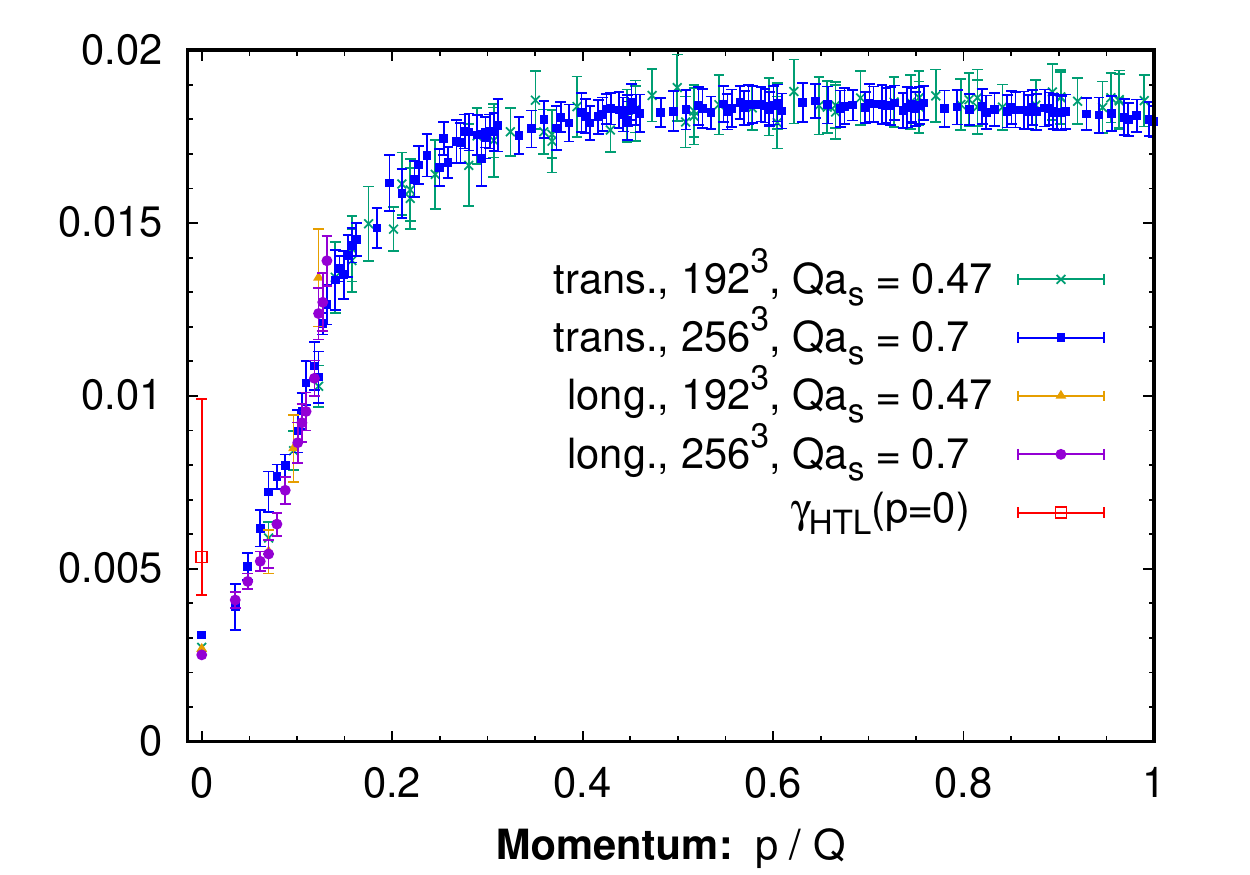}
 \includegraphics[width=0.45\textwidth]{\pToFigs/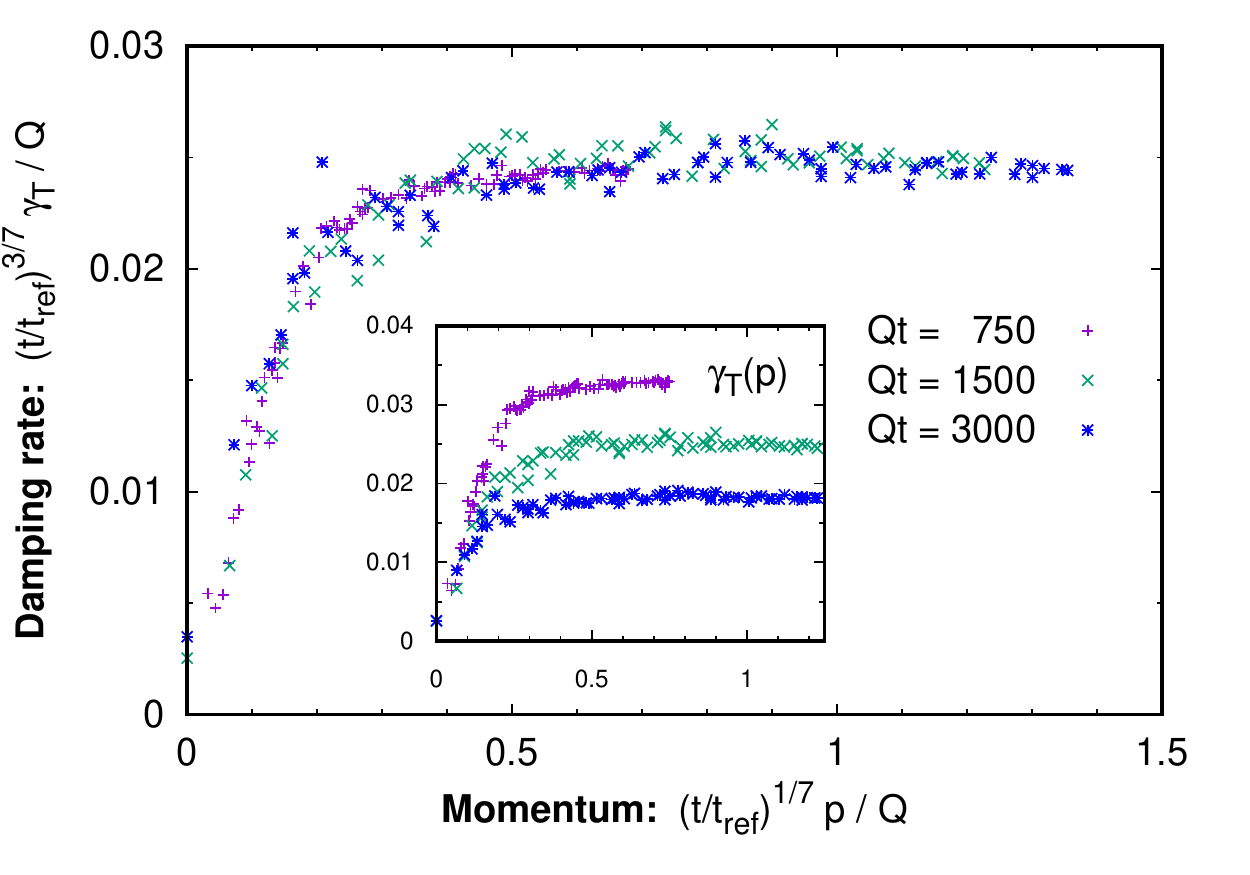}
 \caption{Damping rates $\gamma_\alpha(t,p)/Q$ of an isotropic 3+1D plasma in the self-similar regime as a function of momentum. 
 {\em Left:} For transverse and longitudinal polarizations ($\alpha=T,L$) and different discretizations at fixed time $Q t = 1500$. 
 {\em Right:} Transverse damping rate $\gamma_T(t,p)$ rescaled by $(t/t_{\text{ref}})^{3/7}$ as a function of rescaled momentum at different times, with $Q t_{\text{ref}} = 1500$. 
 Figures taken from \re\cite{Boguslavski:2018beu}.
 }
 \label{fig:gamma_3D}
\end{figure}

We start with \fig\ref{fig:rho_3D} where we discuss the gluonic spectral function of an isotropic 3+1D plasma at a self-similar attractor \cite{Boguslavski:2018beu}. In the left panel we show the spectral function of transversely polarized gluons $\rho_T$ at different momenta $p$ as functions of relative time $\Delta t \equiv t'-t$ (left) and frequency $\omega$ (right). The spectral function is shown to exhibit two distinguishable features: a narrow Lorentzian quasiparticle peak at the position $\omega_T(p)$ (dispersion relation) with the width $\gamma_T(p)$ (damping rate), and the Landau cut region at frequencies $|\omega| < p$. Perturbative results at leading order (LO) HTL (black dashed lines) describe the main features well. The existence of a generalized fluctuation dissipation relation (FDR) between statistical and spectral correlations 
\be
 \label{eq:gen_FDR}
 \frac{\langle EE \rangle_{\alpha}(t,\omega,p)}{\langle EE \rangle_{\alpha}(t,\Delta t {=} 0,p)} \approx \frac{\dot{\rho}_{\alpha}(t,\omega,p)}{\dot{\rho}_{\alpha}(t,\Delta t {=} 0,p)}\,,
\ee
with $\ddot F \equiv \langle EE \rangle$, $\dot\rho_T(\omega,p) \approx \omega \rho_T(\omega,p)$, $\dot{\rho}_{T}(t,\Delta t {=} 0,p) = 1$ and $\alpha=T,L$ corresponding to transverse and longitudinal polarizations, is shown in the right panel of \fig\ref{fig:rho_3D} using transversely polarised gluons as an example.



The damping rates are extracted from $\dot\rho_\alpha(\omega,p)$ by performing fits with a Lorentzian shape and are shown in \fig\ref{fig:gamma_3D}. One observes in the left panel that $\gamma_\alpha$ agree for both polarizations for momenta below the mass $p \lesssim m \approx 0.15\,Q$, despite having different dispersion relations (see Fig.~12 in \cite{Boguslavski:2018beu}). Comparing the values, one finds a separation of scales 
\be
 \label{eq:gwQ_scaleSep}
 \gamma_\alpha(t,p) \ll \omega_\alpha(t,p) \ll Q\,,
\ee
where $\omega(t,p{=}0) \equiv \wplas \sim m$. 
For comparison, we have included curves from different simulations to show the insensitivity to discretization parameters $N_s$ and $a_s$ of our results. 

The finite peak width is a subleading order effect in HTL and is of order $\gamma_\alpha(t,p) \sim g^2 T_*(t) \sim Q (Q t)^{-3/7}$ in the self-similar regime. The point $\gamma_{\rm HTL}(p{=}0)$ corresponds to the HTL result at $p=0$ \cite{Braaten:1990it}. In the right panel of \fig\ref{fig:gamma_3D}, $\gamma_T$ is rescaled with $g^2 T_*(t)$ and becomes time-independent when plotted as a function of rescaled momentum. This confirms $\gamma_\alpha(t,p) \sim g^2 T_*(t)$ and implies that $\frac{\gamma_\alpha(t,p)}{\omega(t,p{=}0)} \sim (Q t)^{-2/7}$ decreases with time, making the quasiparticle peaks even more narrow.


\begin{figure}
\centering
\includegraphics[width=0.48\textwidth]{\pToFigs/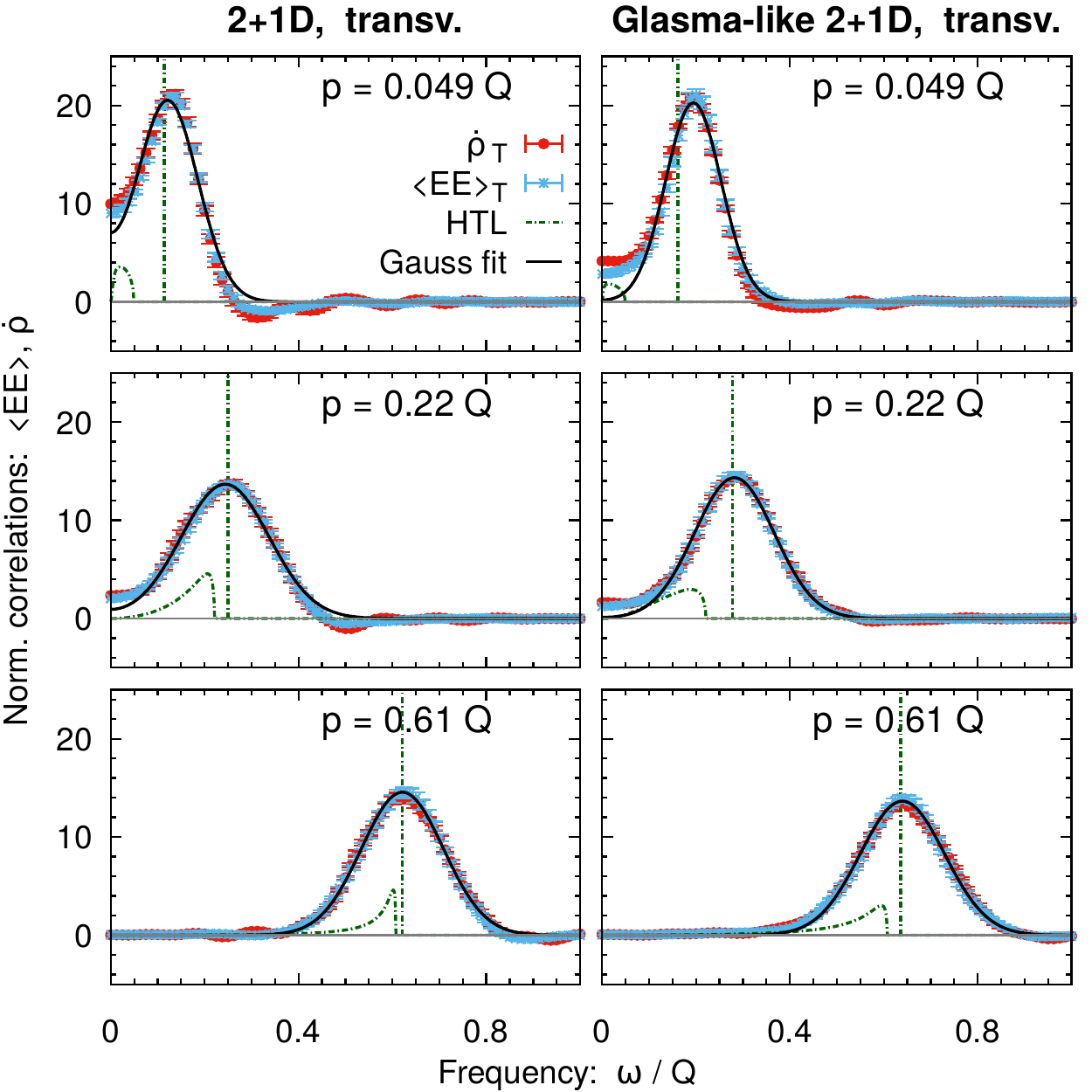}
\includegraphics[width=0.48\textwidth]{\pToFigs/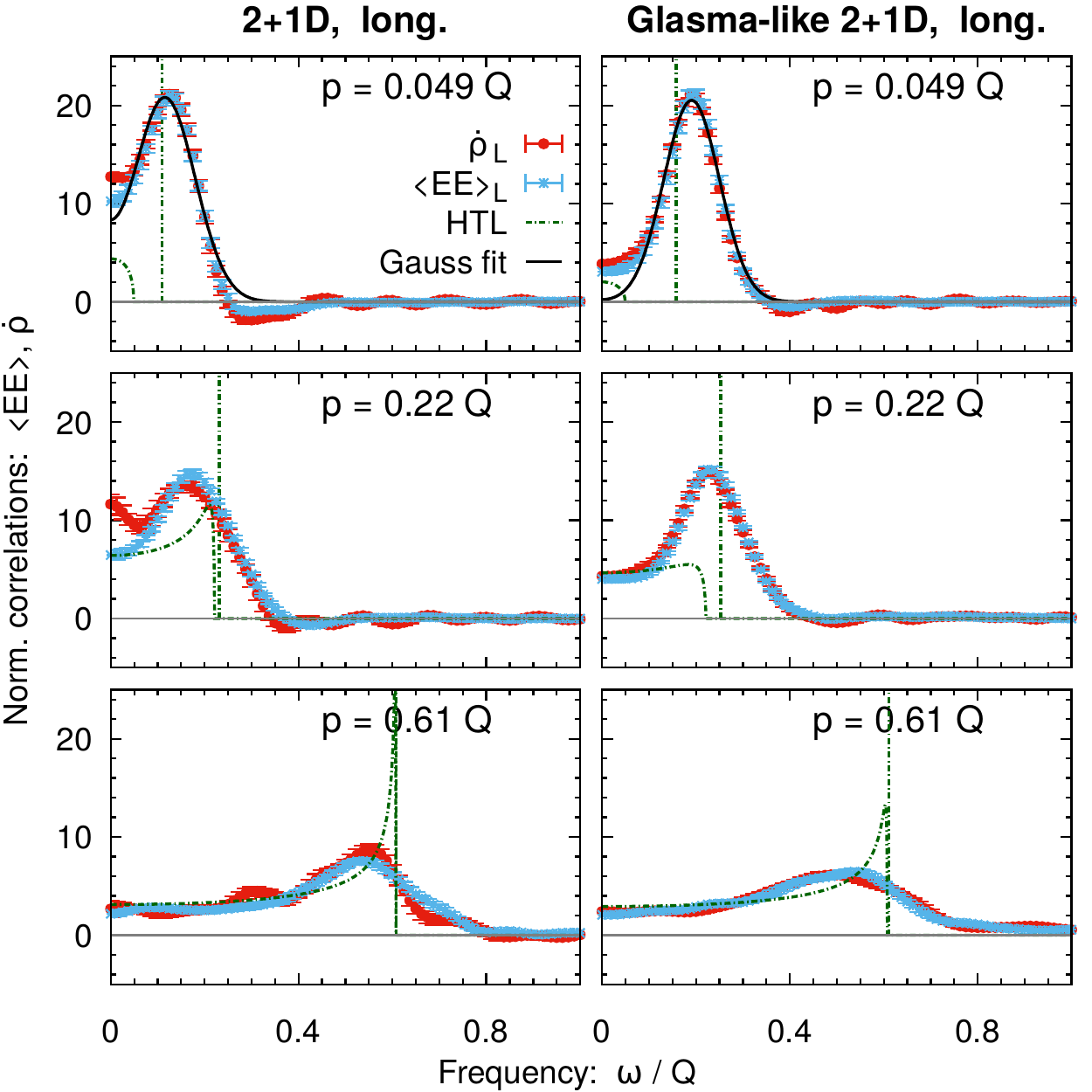}
\caption{
Normalized transverse {\em (left)} and longitudinal {\em (right)} correlations $\dot{\rho}_\alpha(\omega,p)/\dot{\rho}_\alpha(\Delta t {=} 0,p)$ and $\langle EE \rangle_{\alpha}(\omega,p)/\langle EE \rangle_{\alpha}(\Delta t {=} 0,p)$ in 2+1D systems for different momenta. Green dashed lines correspond to HTL results.
Figures taken from \re\cite{Boguslavski:2021buh}.
}
\label{fig:rho_2D}
\end{figure}

\begin{figure}
\centering
\includegraphics[width=0.7\textwidth]{\pToFigs/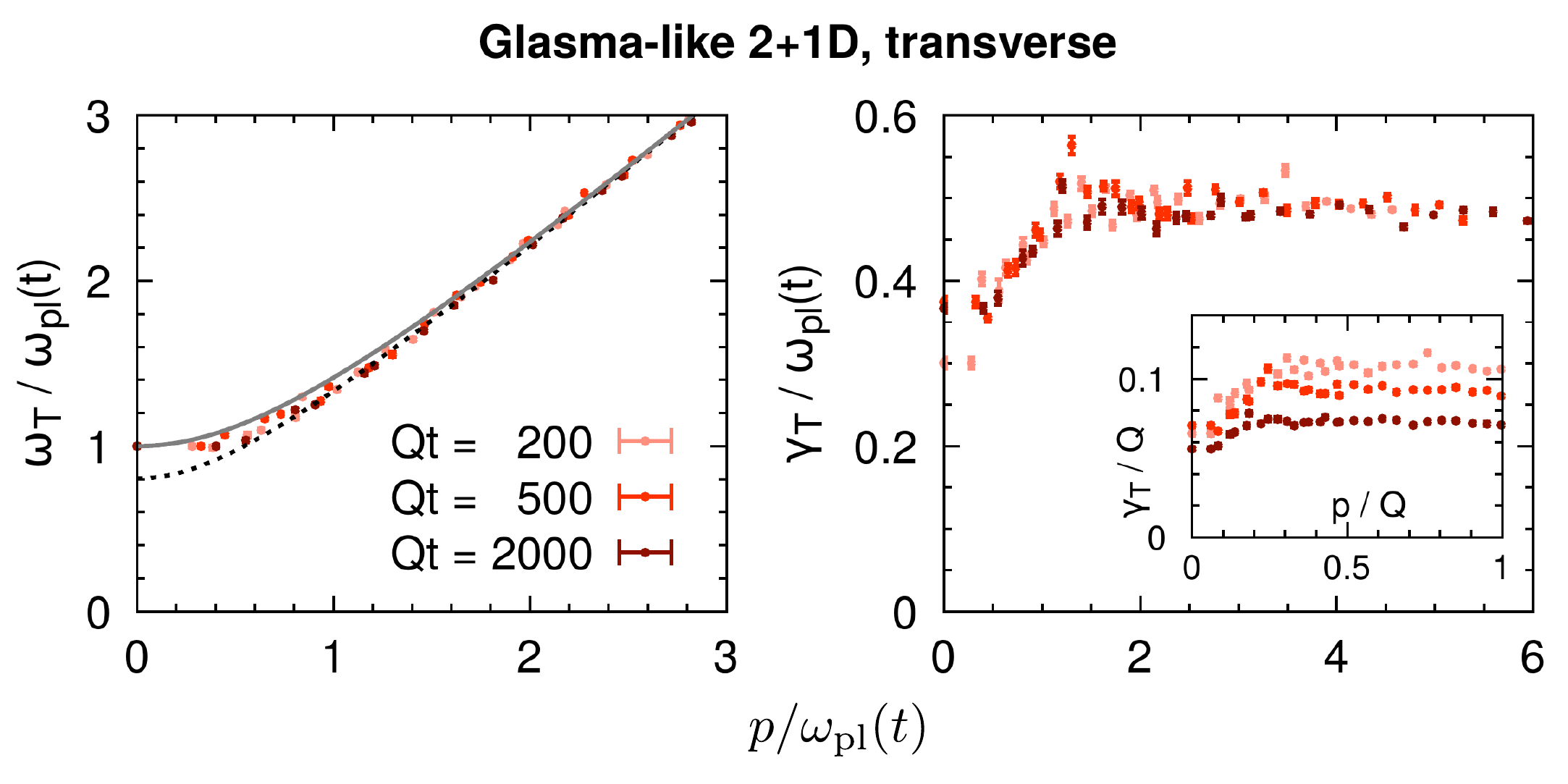}
\caption{
{\em Left:} Dispersion relation $\omega_T(t,p)/\wplas(t)$ and {\em right:} damping rate $\gamma_T(t,p)/\wplas(t)$ of the Glasma-like system at times $Q t = 200$, $500$, $2000$ as a function of $p/\wplas(t)$.
Figures taken from \re\cite{Boguslavski:2021buh}.
}
\label{fig:w_g_2D}
\end{figure}

\begin{figure}
\centering
\includegraphics[width=0.7\textwidth]{\pToFigs/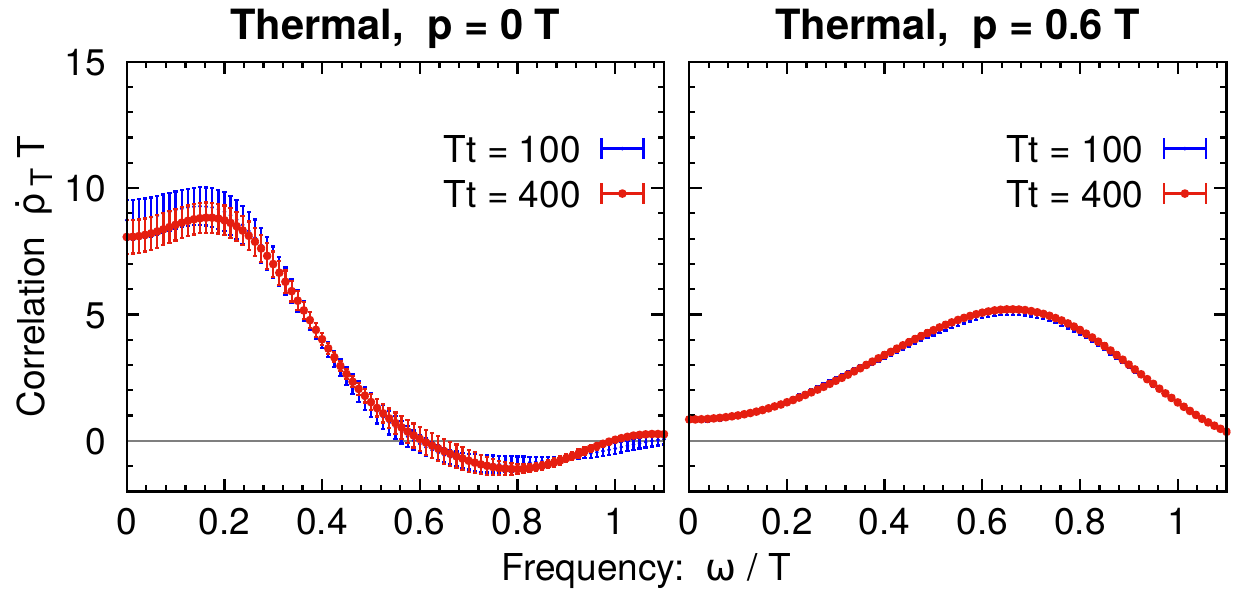}
\caption{
Gluon spectral function in 2+1D classical thermal equilibrium with $f = T / \omega(p)$ (from \cite{Boguslavski:2021buh}).
}
\label{fig:2D_class}
\end{figure}

\newpage
\subsection{Limit of extreme anisotropy: spectral functions in 2+1D plasmas}

\begin{wrapfigure}{l}{0.28\linewidth}
\includegraphics[width=0.28\textwidth]{\pToFigs/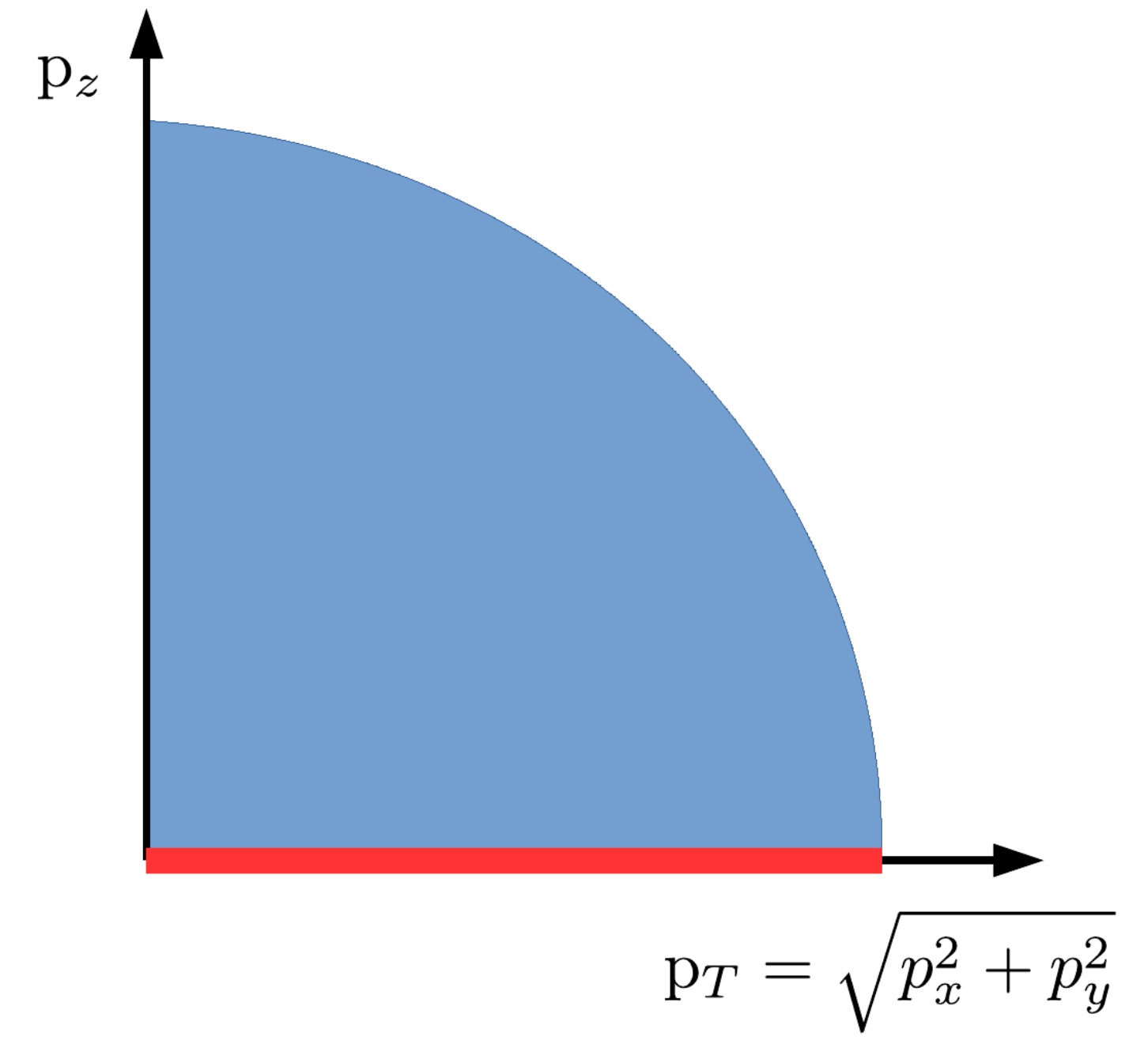}
\caption{{\em Blue:} 3+1D distribution, {\em red:} 2+1D limiting distribution.}
\label{fig:anisotropy}
\end{wrapfigure}

So far, we have discussed the situation of an isotropic 3+1 dimensional system with distribution $f(t,p)$. However, in heavy-ion collisions, the distribution function can be very anisotropic at initial stages, with the limiting case being an effectively 2+1 dimensional system with distribution $f(t,p_T, p_z{=}0)$, as illustrated in \fig\ref{fig:anisotropy} on the left. 
We will again employ Minkowski metric for simplicity, as in the 3+1D case, and extract unequal-time correlation functions from our simulations at the corresponding self-similar attractor (see \re\cite{Boguslavski:2021buh} for a more detailed discussion of our results).

The normalized spectral and statistical correlation functions are shown in \fig\ref{fig:rho_2D} for both transverse and longitudinal polarizations. Both types of correlators lie on top of each other to good accuracy, confirming the validity of the generalized FDR \eqref{eq:gen_FDR} also for 2+1 dimensional systems.

However, there are several differences to the 3+1D case: the resonances are {\em broad peaks} and the correlators remain finite even for $\omega \to 0$. They have a {\em non-Lorentzian shape} that is well approximated by a Gaussian distribution, as shown in \fig\ref{fig:rho_2D}. The green dashed lines correspond to the {\em HTL results} that would be expected in 2+1D, but {\em fail to reproduce} our nonperturbative simulation results. The latter is not too surprising since, as argued already in \re\cite{Boguslavski:2019fsb}, 2+1 dimensional gauge systems lack the scale separation that is required for a hard loop approximation underlying HTL calculations. 

More concretely, we show in \fig\ref{fig:w_g_2D} the transverse dispersion relation and damping rate of the 2+1D Glasma-like system at different times. Rescaling all dimensionful quantities by the plasmon mass $\wplas(t)$, one finds that both become time-independent. This leads to 
\bea
 \gamma(t,p) &\sim& \wplas(t)\,, \\
 \gamma(t,p) &\sim& \omega(t,p) \qquad \text{for} \quad p\lesssim \wplas\,.
\eea
Due to their short lifetime, this implies that no quasiparticles exist for $p \lesssim \wplas$. Note that these statements remain true also at later times, in sharp contrast to the scale separation we observed for isotropic 3+1D gluonic plasmas in \eq\eqref{eq:gwQ_scaleSep}.


We find the same properties also in classical thermal equilibrium, as shown in \fig\ref{fig:2D_class}. Running simulations with a classical thermal distribution $f(p) \approx \frac{T}{\omega(p)}$, we observe that $\rho(\omega,p)$ behaves qualitatively similarly to our far-from-equilibrium case above: one finds broad gluonic excitations of non-Lorentzian shape with $\gamma(p) \sim \wplas$, HTL provides a poor description, and for $\omega \to 0$ the spectral function $\dot{\rho}_T = \omega \rho_T$ remains finite at low $p$. 

This indicates that the extracted properties in 2+1D theories are not special to the non-equilibrium self-similar regime but also emerge in the time-translation invariant classical thermal equilibrium, and may be even found in more general 2+1 dimensional states.

\section{Conclusion}

In this work we have presented a non-perturbative tool developed in \cite{Boguslavski:2018beu}
to extract gluonic spectral functions in highly occupied non-Abelian plasmas far from equilibrium in classical-statistical lattice simulations, and that we have applied to 3+1 and (effectively) 2+1 dimensional systems at classical self-similar attractors (see \res\cite{Boguslavski:2021buh,Boguslavski:2019fsb,Boguslavski:2018beu} for details). 

In general, we find that even far from equilibrium there exists a generalized fluctuation-dissipation relation that relates spectral and statistical correlation functions. However, there are crucial differences between the studied systems. On the one hand, the gluonic spectral function of an isotropic 3+1 dimensional plasma is well described by HTL perturbation theory. Our method allowed us to determine the damping rates of transverse and longitudinal quasiparticle excitations as functions of momentum from first principles. These were much smaller than the dispersion relation, showing that quasiparticles are long-lived and a perturbative description is, indeed, justified. 

In contrast, the gluonic spectral functions of 2+1 dimensional plasmas agree poorly with HTL predictions. They exhibit broad non-Lorentzian excitation peaks with a width of the order of the plasmon mass. This is observed for different times and even in classical thermal equilibrium, indicating that these properties are not special to the classical self-similar attractor, but may be even found in more general 2+1 dimensional states. As a consequence of the broad peaks, momentum modes up to the mass scale have only short-lived excitations and have to be treated non-perturbatively. 

These results are relevant for the description of gluonic excitations at low momenta in heavy-ion collisions and for anisotropic HTL perturbation theory. They show that 2+1D plasmas have genuinely nonperturbative properties. 
Since the effectively 2+1D system stems from an extreme momentum anisotropy of the plasma, our results suggest that there may be significant nonperturbative corrections present also in systems with finite but large anisotropy, which are relevant to phenomenological applications in heavy-ion collisions. 

To approach these questions in more detail, the simulations will be extended to expanding and anisotropic plasmas to study what may be lacking in the current kinetic \cite{Arnold:2002zm,Kurkela:2018wud,Fu:2021jhl} and HTL descriptions. Moreover, we study how nonperturbative properties in correlation functions affect transport coefficients, which has been started in \re\cite{Boguslavski:2020tqz}.


\begin{acknowledgments}
  The author is grateful to A.~Kurkela, T.~Lappi and J.~Peuron for collaboration on different projects presented here. He would also like to thank the organizers for the interesting conference.
  This research was funded in part by the Austrian Science Fund (FWF) project P 34455-N. 
  The author wishes to acknowledge the CSC - IT Center for Science, Finland, and the Vienna Scientific Cluster (VSC) for computational resources. 
\end{acknowledgments}



\end{document}